\documentstyle[preprint,aps]{revtex}

\tightenlines

\newcommand{\vect}[1]{\mbox{\boldmath $#1$}}

\newcommand{\lsim}[1]{
\setlength{\unitlength}{12pt}
\begin{picture}(1.4,1.)
\put(.7,-0.3){\makebox(0.0,1.)[t]{$<$}}
\put(.7,-0.3){\makebox(0.0,1.)[b]{$\sim$}}
\end{picture}#1}
\begin{document}
\draft

\title{Medium effects in string-dilaton-induced neutrino oscillations}

\author{R.Horvat \\
   ``Rudjer Bo\v skovi\' c'' Institute, P.O.Box 1016, 10001 Zagreb,
Croatia}

\maketitle

\begin{abstract}

We consider the unconventional way to interpret the current data on
solar neutrino oscillations as derived recently by Halprin and Leung from a
string model based on the existence of the string dilaton field which
remains massless in the low-energy world. The equivalence principle
violation entailed by the existence of a massless dilaton may then produce
neutrino oscillations even for neutrinos that are degenerate in mass. Here we
calculate the medium-induced mass squared difference for solar and
atmospheric neutrinos, which is due to their coherent interactions with the 
cosmic neutrino background and with solar plasma constituents. We show that 
this difference can naturally be large enough to satisfy the known 
experimental limits on the oscillatory solutions of the solar as well as 
atmospheric neutrino problem.
\end{abstract}

%\newpage

   Among the three related  ways to produce neutrino oscillations for massless
or degenerate-in-mass neutrinos \cite{1}, two of them are based on the
equivalence principle violation by neutrinos, although in very different
ways. In one of them \cite{2}, the neutrino mixing was generated by a
violation of the equivalence principle induced by a breakdown of
universality in the gravitational coupling strength between the conventional
spin-2 particles and the neutrinos. In the other approach \cite{3}, Halprin
and Leung made use of a scenario in string theory proposed recently by
Damour and Polyakov \cite{4}, in which the string dilaton field remains
massless and violates the equivalence principle. It has been stated \cite{4}
that a violation of the equivalence principle is now entailed solely by a
massless dilaton, with the neutrino mixing being consequently generated due
to a breakdown of universality in the coupling strength between the spin-0
particles  and the neutrinos \cite{3}. In both examples listed above,
neutrino oscillations will take place even for neutrinos that are degenerate
in mass (or even for massless neutrinos in the first example).

   In the analysis \cite{3} based on a dilaton violation of the
equivalence principle, the largest contribution to a local value of the
neutrino effective mass squared difference  in an external
gravitational field
was taken to come from a very massive system, called Great Attractor,
causing the large regions of about 100 Mpc of our local neighborhood to
move towards it with peculiar velocities of about 600 km/sec.
 It has been shown by Halprin and Leung \cite{3} that by
taking a degenerate  mass at the experimental limit on
$m_{\nu_{e}}$, one is able to obtain an effective mass difference for solar 
neutrinos large enough to satisfy the known experimental limits for the
vacuum oscillation solution \cite{5} of the solar neutrino problem. On the
contrary, the effective mass difference is too small to satisfy the MSW
solution \cite{6}, unless one is willing to deal with unnaturally large values
of  neutrino-dilaton couplings (on which no severe constraints exist). Here
we demonstrate how a matter background via  coherent interactions can also
induce an extra mass for neutrinos, which can  naturally be large enough to
provide even the MSW solution. We show that the main contribution depends
only on the ratio of the matter-dilaton coupling constants, making thereby
its value practically independent of any constraint. In the following we 
study the influence of the cosmic neutrino background as well as  
solar plasma on  neutrino masses. 

   We begin our discussion  with a possibility discussed in the scenario of
Damour and Polyakov:That the dilaton remains massless and that the
matter-dilaton coupling constants are species dependent which means nothing
other than a violation of the equivalence principle. A spin-0 exchange
contribution to the static gravitational energy can be described by the
following lagrangian density \cite{3}
\begin{equation}
{\cal L}= \sqrt{4\pi G_{N}}\,\alpha_{A}m_{A}\bar{\psi}_{A}\psi_{A}\phi \;\;\; ,
\label{form1}
\end{equation}
where $\psi_{A}$ and $\phi$ is a matter field of type $A$ and the dilaton
field, respectively. Here $G_{N}$ is the Newton's gravitational constant and
$\alpha_{A}$ is the present strength \cite{4} of the coupling of the dilaton
to $A$-type particles. From Eq.(1), the interaction potential energy between
particle $A$ and particle $B$ contains the $\alpha$-dependent piece, beside
the usual spin-2 exchange contribution,
\begin{equation}
V(r)=-G_{N} m_{A} m_{B} (1+\alpha_{A} \alpha_{B})/r \;\;\; .
\label{form2}
\end{equation}   
In \cite{3} the induced neutrino mass in an external gravitational field was
given essentially by formula (2), where the label $A$ refers to a neutrino
and label $B$ refers to a static matter distribution whose potential was
taken to be dominated by the Great Attractor, with a value around 
$3 \times 10^{-5}$ \cite{7}. There is also a limit on $\alpha_{B}$ coming
from conventional solar-system gravity experiments, which is 
$\alpha_{B}< 10^{-3}$ \cite{8}. This constrains the induced mass
additionally, as one assumes $\alpha_{\nu} \lsim 10^{-3}$.  

    Our main observation here is that the induced mass for neutrinos,
arising from  coherent interactions of $\nu$'s with the background matter,
depends only on the ratio of $\alpha$'s but not on their absolute values,
thereby making its value much less constrained than for the 
static-matter-distribution
case. Let us concentrate first on the background neutrinos at an effective
temperature $T_{\nu} \simeq 1.9\; K$. In the framework of thermal field theory,
it can be shown that the major contribution to the induced neutrino mass
comes from a tadpole diagram (where a massless dilaton  carries no
four momentum and with cosmological neutrinos running inside the
medium-induced loop). Note that the tadpole diagram is a constant
independent of the external neutrino momentum. In the real-time version
\cite{9} of thermal field theory, the computation of $m^{ind}$ is
straightforward and reads
\begin{equation}
  m^{ind}_{l} = -8 \pi G_{N} \frac{\alpha_{l} \, \alpha_{h} \, m_{l} \,
  m^{2}_{h}}{m_{el}^{2}} \int \frac{d^{3}k}{(2 \pi)^{3} E_{k}} \, n_{F}(k)
\;\;\; ,
\label{form3}
\end{equation}
where
\begin{equation}
n_{F} (k) = \frac{1}{e^{\beta k} + 1} 
\label{form4}
\end{equation}
is the distribution function  of a  relativistic neutrino,
with $1/\beta \equiv T$ and $E_{k}= \sqrt{k^{2} + m^{2}_{h}}$. Also, in
Eq.(3) and (4), $k\equiv|\vect{k}|$. As for Eq.(3), several remarks are
in order. First, we compute $m^{ind}$ for a light neutrino  ($l$)
since we are interested in two-neutrino mixing between light neutrinos (say
$\nu_{e}$ and $\nu_{\mu}$), which is to solve the solar neutrino problem.
Secondly, we assume, for simplicity, that $m^{ind}$ in Eq.(3) is dominated
by the heaviest neutrino from the background (having a mass $m_{h}$ here), as
one might naively expect from (1). If the heaviest neutrinos are so massive
that they are nonrelativistic today, they still should be described by the
relativistic distribution function of the form (4), as their total number
density is fixed at about $100\; \mbox{\rm cm}^{-3}$. These considerations bring us into
the realm of out of equilibrium thermal field theories as Eq.(3) is
obtained with the aid of (11) component of the real-time neutrino propagator 
of the form \footnote{The problem with ill-defined pinch singularities, which 
is characteristic for out of equilibrium thermal field theories, does not
obtain here.}
\begin{equation}
     S_{11} (k) = ( \not{\!k} + m_{h} )
            \left( \frac{\textstyle 1}{\textstyle k^{2} - m_{h}^{2} + i
\epsilon} + 2 \pi i \,
            n_{F} (k) \, \delta (k^{2} - m_{h}^{2} ) \right) \;\;\; .
\label{form5}
\end{equation}    
Further, in the massless-soft regime, the resummation program developed by
Braaten and Pisarski \cite{10} must be applied, resulting in the use of the
resummed propagator for a massless dilaton in (3). The static limit of the
resummed propagator is then responsible for the appearance of $m_{el}$ in
(3), the analogue of the electric mass in gauge theories. Also, we take the
number of neutrino degrees of freedom to be equal 2 in Eq.(3), because 
for light Dirac masses right-handed neutrinos decouple so early that their 
contributions to the cosmological energy density were diluted relative to 
left-handed neutrinos, by the later annihilation of particle species. 
Finally, since the cosmic neutrino background is likely to be 
$CP$-symmetric, the chemical potential is assumed to vanish for all
neutrinos in Eqs.(3) and (4). 

   The gravitational screening $m_{el}$ is defined by taking the static
limit of the temperature dependent one loop self-energy diagram for the
dilaton; that is, $\Pi_{d}(q_{0}=0, \vect{q} \rightarrow \vect{0} )$. 
If all neutrinos are nearly massless even the cosmic neutrino background remains
relativistic today, and by explicit calculation we find in this case that
\begin{equation}
    m_{el}^{2} \simeq \frac{\pi}{3}\,G_{N}(\alpha_{h}^{2} m_{h}^{2})T^{2} 
    \;\;\;.
\label{form6}
\end{equation} 
If, on the other hand, the loop neutrinos have nonrelativistic velocities
today, then they have much stronger impact on dispersion, giving, 
\begin{equation}
     m_{el}^{2}\simeq -\frac{4 \ln 2}{\pi}\,G_{N} (\alpha_{h}^{2} m_{h}^{3})T
\;\;\;.
\label{form7}
\end{equation}
Note the occurrence of $m_{el}^{2}<0$ in (7), or a ``thermal tachyon''. If
the electric mass were negative, the effective propagator would have a tachyon
and the theory would be thermodynamically unstable. For scalar or gauge
theory in four dimensions, the electric mass at one loop order is always found to
be positive. This is reasonable, as these theories are thermodynamically
stable. However, when a theory has interactions which are universally
attractive, a thermal distribution just tends to collapse upon itself,
causing $m_{el}^{2}$ to become negative (see, e.g., the example of graviton
vacuum polarization in Ref.\cite{11}). Indeed, $m_{el}^{-1}$ is nothing other
than an extra contribution to the usual Jean's length - beyond which neutrinos
do not erase fluctuations. On the contrary, $m_{el}^{2}$ from Eq.(6) is
positive; as we are now in the relativistic regime, $m_{\nu}<<T$, and
because neutrino dominance occurs at a temperature $T$ of order $T\; \sim
\;m_{\nu}$, the free streaming of relativistic particles erases fluctuations
on all scales within the horizon. One can notice that the
``attractiveness'' of the interaction (1) does not show up in the
calculation of $m_{el}^{2}$ before $T\; \sim\;m_{\nu}$.

   Now, by combining Eqs.(3), (6), and (7), we end up with the final result
for the induced mass

\begin{eqnarray}
m_{l}^{ind}\;\simeq\;-\;\left(\frac{\textstyle \alpha_{l}}{\textstyle \alpha_{h}} \right)\;m_{l}
&
~~~~~~~~~~(T>>m_{h})~,
\label{form8}
\\ \nonumber
~
\\
\vspace*{5mm}
m_{l}^{ind}\;\simeq\;2.6\;\left( \frac{\textstyle \alpha_{l}}{\textstyle \alpha_{h}} \right)
\;\left( \frac{\textstyle T^{2}}{\textstyle m_{h}^{2}} \right)
\;m_{l}
&
~~~~~~~~~~(T<<m_{h})~.
%\label{form9}
\end{eqnarray}

Note the interesting case of cosmological relic neutrinos which would have
remained undisturbed until the present time, Eq.(8), where it is just the 
violation of the equivalence principle in the neutrino sector that is
responsible for the prevention of cancellation between the vacuum and the
induced mass (apart from a small correction of order of $m^{4}/T^{4}$ not
displayed in Eq.(8)). On the contrary, we can see that, apart from a different
sign, $m_{l}^{ind}$ for the case of gravitational instability, Eq.(9), is
always reduced by a factor $T^{2}/m_{h}^{2}$ with respect to the former
case. It is of utmost importance to notice that it is only the ratio of
$\alpha$'s that shows up in Eqs.(8) and (9), on which practically no
constraints exist (note also that the ratio can be both less and larger than
unity).

   A violation of the equivalence principle will be  most noticeably if
the oscillatory neutrinos are completely degenerate but not massless. The
degeneracy can be protected by a presumed global inter-family (flavor or
horizontal) symmetry of leptons. Then,
\begin{equation}
   m_{l_{i}}^{*}=\;m_{l_{i}}\;+\;m_{l_{i}}^{ind}\;\;\;\;\;\;,
\label{form10}
\end{equation}
with $m_{l_{1}}=m_{l_{2}}=m$. Concerning ourself with the case where the mass
and the gravitational eigenstates (that is, those defined through Eq.(1))
are identical, the expression for flavor survival probability (for say two
flavors with mixing angle $\theta $) as a function of distance $L$ turns out to
be
\begin{equation}
   P(\nu_{e} \rightarrow \nu_{e})=1-{\sin ^{2}(2 \theta )}\;{\sin ^{2}\left(
\frac{L\Delta m^{*2}}{4E} \right)} \;\;\;\;\;\;,
\label{form11}
\end{equation}
where $\Delta m^{*2} \equiv m_{l_{2}}^{*2}-m_{l_{1}}^{*2}$. Taking, for
instance, the case given by Eq.(9) to be the relevant one, we find
\begin{equation}
  \Delta m^{2} \simeq 5.2\; m^{2} \left( \frac{T^{2}}{m_{h}^{2}} \right) 
\frac{\Delta\alpha}{\alpha_{h}}\;\;\;\;\;\;,
\label{form12}
\end{equation} 
where only terms to the first order in $T^{2}/m_{h}^{2}$ are kept,
and $\Delta \alpha=\alpha_{2}-\alpha_{1}$. The observed deficit of solar
$\nu_{e}$'s can be explained by two-neutrino mixing with $\Delta m^{2}\sim
10^{-5} \;\mbox{\rm eV}^{2}$ in the case of MSW transitions or with 
$\Delta m^{2}\sim10^{-10} \,\mbox{\rm eV}^{2}$ in the case of vacuum 
transitions. If, for the purpose of illustration, we take 
$m_{l} \sim 1\; \mbox{\rm eV}$; $m_{h} \sim 10\; \mbox{\rm eV}$, then we need 
$\mid \Delta \alpha / \alpha_{h} \mid \sim 10^{-1}$ to fit the vacuum 
oscillation solution and $\mid \Delta \alpha / \alpha_{h} \mid \sim 10^{4}$ 
to fit the MSW solution. (Notice, however, that it is actually a
demand on $\mid \Delta \alpha / \alpha_{l} \mid$ in the latter case as our
assumption of the heaviest-neutrino dominance breaks down in that case.) If the 
halo of our Galaxy were partially contributed by eV-neutrinos (most at the
level of a few percent, see Ref. \cite{12}), then $T_{\nu} \simeq 50\,K$, and thus 
$\mid \Delta \alpha / \alpha_{h} \mid \sim 10^{-4}$ or 
$\mid \Delta \alpha / \alpha_{h} \mid \sim 10$, respectively. We do not
bother here to  display a formula for the case when all three neutrinos are 
degenerate in mass (for instance, with $\sum_{i} m_{\nu_{i}}=5\,\mbox{\rm eV}$
 as preferred in the theory of formation of cosmic structure in a 
critical-density universe, see Ref. \cite{13}) as it is  again given in a 
form of the ratio of charges similar to that in Eqs.(8) and (9). 

    The recent data from the Super-Kamiokande collaboration \cite{14}
for the up-down asymmetry in the atmospheric muon-neutrino flux provides a
strong evidence for oscillations of muon neutrinos into nearly maximally
mixed tau or sterile neutrinos. In the original scenario \cite{15}, which
aimed to solve all the neutrino anomalies by introducing a fourth neutrino,
the atmospheric neutrino problem is solved via the $\nu_{\mu}-\nu_{\tau}$
oscillation with maximal mixing and nearly degenerate-in-mass neutrinos. Hence, 
this fits naturally with the above oscillation mechanism. It follows from
Eq.(12) that $\Delta m^{2}$ is actually independent of the neutrino mass for
two-neutrino mixing between heavy neutrinos ($\nu_{\mu}$ and $\nu_{\tau}$). The
atmospheric neutrino data requires $\Delta m^{2}\sim10^{-3}-10^{-2}
\,\mbox{\rm eV}^{2}$. For a backgroud consisting of halo neutrinos one
therefore needs $\mid \Delta \alpha / \alpha_{h} \mid \sim 10-10^{2}$ in
order to fit the oscillatory solution of the atmospheric neutrino problem. 

    One is, however, under an obligation to mention some ambiguities
inherent to the calculations just outlined above. It is to be noted that all
macroscopic quantities, introduced by coherent interactions by means of
averaging over the background particles (all of them are certain functions
of the temperature for thermal equilibrium), do not necessarily refer to local
quantities (here by local we mean with respect to the position of a
neutrino). This occurs because our interaction is long-ranged. In matter,
however, we expect to have a screened interaction whose range is cutoff at
distances of order of $m_{el}^{-1}$ \cite{16}. Still, $m_{el}^{-1}$ cannot be
quantified exactly as it depends on the unknown $\alpha$'s. In any case, it
seems that one has to give up the simplicity of the homogeneous background
and deal with inhomogeneous multi-component equilibria. In view of the above
uncertainties \footnote{Similar uncertainties for values of the
gravitational potential at various positions from various sources were
discussed in first Ref.\cite{2}.}, and the fact that MSW transitions take place
in the body of the Sun, it is also necessary to calculate $m^{ind}$ coming
from the solar plasma itself. In this case we find that
\begin{equation}
  m_{l}^{ind}\;\simeq\;\left( \frac{\alpha_{l}}{\alpha_{p}} \right)
\;\left( \frac{T}{m_{p}} \right) \; m_{l}\;\;\;\;,
\label{form13}
\end{equation}
where $T$ is the solar temperature and $m_{p}$ is the proton mass. Comparing
this with our previous results, one can see that both the ratio of $\alpha$'s and
the background-dependent part are now different  to that given by Eqs.(8) and (9). As
$\alpha_{p}$ is probably constrained (recall our discussion at the
beginning), a ratio $\alpha_{l}/\alpha_{p}$ is still undetermined and can 
easily be tuned to fit the MSW solution. If the net contribution to the
induced mass were dominated by Eq.(13), then for $m_{l} \sim 1\; \mbox{\rm
eV}$ we need
$\mid \Delta \alpha / \alpha_{p} \mid \sim 10-10^{4}$. Notice, however, that
a more sophisticated treatment is needed in this case, since the mass squared 
difference is now temperature dependent and would consequently influence the
resonant condition for neutrinos propagating through the Sun.  
  
   In conclusion, we have calculated an effective mass for neutrinos from a 
 matter background by considering a nonuniversal scalar gravitational
interaction where the spin-0 field is exactly massless in the vacuum. By
using the real-time approach of thermal field theory and considering only
those graphs having a tadpole topology, we have calculated the effective
mass difference for solar neutrinos induced by the cosmic neutrino background
and  solar plasma by means of coherent interactions. In comparison with
the induced mass by the gravitational potential from a static matter
distribution, we have found a clear distinction. While this latter mass
difference depends on the difference between the coupling constants of the
two neutrino species, the former always involves the ratio of coupling
constants, which, being totally unconstrained, can easily be tuned to fit
the  most popular solutions of the solar as well as atmospheric neutrino
deficit. \newline

{\bf Acknowledgments. } The author acknowledges the support of the Croatian
Ministry of Science and
Technology under the contract 1 -- 03 -- 068.

\end{document}